\documentclass[aps,pre,onecolumn,notitlepage,groupedaddress]{revtex4-1}

\usepackage{color}
\usepackage{amsmath}
\usepackage[pdftex]{graphicx}
\usepackage{subfigure}
\usepackage{lipsum}
\usepackage{lineno}
\usepackage{todonotes}
\usepackage[normalem]{ulem}

\begin{document}

\title{Abrupt transition due to non-local cascade propagation in multiplex systems}

\author{Oriol Artime}
\author{Manlio De Domenico}
\affiliation{CoMuNe Lab, Fondazione Bruno Kessler, Via Sommarive 18, 38123 Povo (TN), Italy}
\date{\today}

\begin{abstract}
Multilayer systems are coupled networks characterized by different contexts (layers) of interaction and have gained much attention recently due to their suitability to describe a broad spectrum of empirical complex systems. They are very fragile to percolation and first-neighbor failure propagation, but little is known about how they respond to non-local disruptions, as it occurs in failures induced by flow redistribution, for example. Acknowledging that many socio-technical and biological systems sustain a flow of some physical quantity, such as energy or information, across the their components, it becomes crucial to understand when the flow redistribution can cause global cascades of failures in order to design robust systems,to increase their resilience or to learn how to efficiently dismantle them. In this paper we study the impact that different multiplex topological features have on the robustness of the system when subjected to non-local cascade propagation. We first numerically demonstrate that this dynamics has a critical value at which a small initial perturbation effectively dismantles the entire network, and that the transition appears abruptly. Then we identify that the excess of flow caused by a failure is, in general, more homogeneously distributed the networks in which the average distance between nodes is small.Using this information we find that aggregated versions of multiplex networks tend to overestimate robustness, even though to make the system more robust can be achieved by increasing the number of layers. Our predictions are confirmed by simulated cascading failures in areal multilayer system.
\end{abstract}

\maketitle

\section{Introduction}

Power grids, trophic relations, online micro-blogging services or scientific collaborations: complex networks are ubiquitous in nature~\cite{barabasi2003linked,newman2018networks}. Since the nascence of modern network science more than two decades ago, efforts have been made towards finding common characteristics among complex systems of different sort. One of the most salient examples is the robust-yet-fragile property~\cite{albert2000error}: it suffices to have a large variance in the distribution of the number of connections to observe that networks are very good at handling random failures but are extremely weak when facing localized attacks~\cite{cohen2000resilience, cohen2001breakdown}. The study of network robustness sheds light on the characteristics that play a central role on maintaining the proper functioning of these networks and how their different properties are altered when some components are corrupted. This has been a topic of paramount importance in the research agenda of network science due to its conceptual, modelling challenge~\cite{callaway2000network, newman2002spread, holme2002attack, motter2002cascade, motter2004cascade, shao2015percolation, gao2016universal, zhao2004attack, guan2018general} and its wide range of applications~\cite{kinney2005modeling, buldyrev2010catastrophic, yang2017small, may2008complex, heiberger2014stock, dubois2016linking, artime2020effectiveness}. 

From the modeling side, the simplest conceptual way to consider the problem of network robustness is to disregard any intrinsic dynamics in the nodes or in the links, and look for the geometrical properties of the remaining network after the failure or the attack from the viewpoint of classical percolation \cite{stauffer1991introduction}, such as the size of the giant connected component, the cluster size distribution and the nature of the phase transition. The next level of complexity includes the spreading of failures across the network, for example, by letting a node collapse if its fraction of failed neighbors is higher than a threshold~\cite{watts2002simple}, by mapping the propagation to a branching process~\cite{brummitt2012suppressing, lee2004sandpile}, or even by combining different rules~\cite{yu2016system}. All these models share the limiting feature that the failure propagates via first-neighbors connections. This allows mathematical tractability but does not capture the non-local propagation observed in some real-world cascade failures, e.g. in the 1996 disturbance of the Western Systems Coordinating Council (WSCC) system~\cite{NERC2002system}, in the 2003 blackout in northeast USA \cite{nerc2004technical} or in the air-traffic disruption due to the eruption of the Icelandic volcano Eyjafjallaj{\"o}kull \cite{eurocontrol2010ashcloud, wikipedia}. The family of models that offers a plausible explanation for this non-local chain of failures is the one that considers the flow of some quantity across the network~\cite{motter2017unfolding, motter2002cascade, motter2004cascade, crucitti2004model, moussawi2017limits, waniek2020traffic}, such as the electrical energy in the power grid or the number of passengers between public transport stations. When the structure of the network is perturbed, for instance by removing a small subset of nodes, an overall flow redistribution is observed and subsequent failures might occur, not necessarily close to the previous failure.

The flow redistribution can occur in various ways depending on the characteristics of each system. Embracing this heterogeneity, several types of models have been proposed in the past. As it is usual in mechanistic models of statistical physics, we have simple, toy models that seek plausible physical mechanisms that are able to explain the most remarkable features of the empirical observations, providing qualitative explanations of emergent behaviors via well-isolated cause-effects relations~\cite{baxter2016exactly, castellano2009statistical, artime2019herding}. One of the most paradigmatic examples of this family of models is the one proposed by Motter and Lai~\cite{motter2002cascade}, where nodes have a binary state --- operative or failed --- and the flow is completely determined by the topology. Since our goal is not the accurate reproduction of behavior of a particular system but the qualitative understanding of the role played by several multiplex properties in the propagation of non-local cascades, throughout the article we adopt the aforementioned framework. Further details of our model are given in Section~\ref{sec:model}. It is worth mentioning, however, that there is a plethora of models proposed to describe the phenomenon of non-local cascades, with different level of abstraction and whose suitability depends on the particular scientific question one wants to address, see for example~\cite{dobson2007complex, pahwa2014abruptness, ruan2007spatial, zapperi1997first, simonsen2008transient}. The general tendency is that the more details introduced in a model, the better it can reproduce empirical observations but at the expense of less transparent interpretations of the role played by each of its parameters.

The topological substrates on which we consider the non-local cascade propagation driven by load redistribution are multiplex networks~\cite{de2013mathematical, kivela2014multilayer}. These are graphs composed by a same set of nodes interacting through different contexts, the layers. The multilayer framework is particularly useful to model systems with mutual interdependencies, a real-world feature, and they display a very rich phenomenology that cannot be otherwise observed in aggregated, single-layer networks~\cite{gomez2012evolution, diakonova2016irreducibility, de2013mathematical, de2014navigability, artime2017joint}. Regarding the issue of the robustness in multilayer networks, the same division mentioned above is valid: percolation analyses~\cite{min2014network, bianconi2017epidemic, radicchi2017redundant} and first-neighbor cascade propagation models are abundant~\cite{buldyrev2010catastrophic, brummitt2012multiplexity, lee2014threshold, reis2014avoiding, turalska2019cascading}, while non-local cascade propagation is a promising, but still largely unexplored topic, except for a few exceptions~\cite{zhou2017overload}. Similarly, the characterization of non-local failures in real multilayer systems is certainly interesting to understand how failures can propagate within and across layers. This is indeed possible: for instance, in the examples of real systems displaying non-local cascades given above (power grid and European Air Transportation network) accept a description as multilayer structures~\cite{nardelli2014models, cardillo2013emergence}. To the best of our knowledge this has not been done yet, although it would be a very nice contribution because it would provide a benchmark to compare with the outcome of the dynamical models.

We work under the assumption that connectivity is a first proxy for functionality, therefore the robustness of the network is related to the size of the largest connected component once the cascade stops. We numerically show that overload cascades can display an abrupt phase transition~\cite{d2019explosive} that separates the regimes in which the initial perturbation completely dismantles the network or leaves it almost intact. The transition is characterized by techniques borrowed from the theory of critical phenomena. Beyond the order of the phase transition, which certainly has implications for network fragility assessments, we also address the question of how the robustness is modified if the multiplexity of a network is disregarded and, instead, the aggregated system is considered. In the same vein, we explain how the number of layers impact the overall robustness. To answer these questions, we present compelling evidence that when the perturbed flow can be more homogeneously redistributed across the network, the tendency of robustness is to increase, and quantify this effect by means of the average path length~\cite{newman2018networks}. To check the validity of our results, we apply our model to the European air transportation multiplex network, finding a good qualitative match.

$ $

\section{Model definition} \label{sec:model}

The model that we use is inspired by the load-capacity model of Ref.~\cite{motter2002cascade}. In our case we deal with a multiplex topology formed by $M$ independent networks ($M \geq 2$ corresponding to the number of layers), each of them composed of $N$ nodes and created from the same degree distribution but with different degree sequences. Besides the intralayer links, every node is one-to-one connected with its replica node in the $M-1$ remaining layers. Each node $i$ in layer $m$ is assigned a load $L_{i}^{m}$ and a capacity $C_{i}^{m}$. Loads are computed as the number of shortest paths crossing each node (the interlayer connections also count in the computation) and vary as the cascade propagates. The capacity is a fixed quantity, proportional to the initial load $L_{i}^{m, ini}$, that is, 
\begin{equation}
    C_{i}^{m} = (1 + \alpha) L_{i}^{m, ini}.
\end{equation}
The parameter $\alpha$ is called the tolerance, and gives an idea of the ability of the nodes to readapt to flow changes. When the load of a node exceeds its capacity, $L_{i}^{m} > C_{i}^{m}$, that node automatically fails by breaking all its connections. Moreover, the replicas of the overloaded node in the other layers also fail by disconnecting themselves from their neighbors. We consider cascades that propagate in a simultaneous manner, that is, at each time step all overloaded nodes fail at once, and then the loads are recalculated. To simplify the analysis, the initial perturbation is performed by deleting the node with highest load, i.e., we are dealing with intentional attacks that assume global knowledge of the network. Despite being single-node attacks, we will see that the consequences can be catastrophic for the integrity of the network.

The load-capacity model by Motter and Lai was originally introduced in single-layer networks and it provided insights on how the initial load distribution and thus, the topology, can influence the robustness of the system. However, the nature of the transition --- continuous or discontinuous --- is a problem that was left unexplored in the original article and has been ignored in other works that followed up. To the best of our knowledge, the only attempt to characterize it is Ref.~\cite{kornbluth2018network}, that finds a first-order transition but for massive initial random attacks, instead of single-node attacks. Here our focus is on the response of multiplex systems to the overload cascades, but, additionally, we shed light on the overlooked problem of the order of the phase transition in monoplexes subjected to single-node targeted attacks. In both cases we find an abrupt transition, evincing that it is the dynamics and not the topology who determines the nature of the phase transition.

So far we have been using the concepts of robustness and fragility in an intuitive way. To be more precise, we employ them following the next criterion. On the one hand, a network experiencing a first-order transition is considered by default more fragile than one experiencing a continuous transition, because the transition is more abrupt, hard to anticipate and of more destructive power. On the other hand, when dealing with the same type of transition, we say that the lower the critical point (critical tolerance $\alpha_c$), the more robust the network is. When critical points are similar, then the faster the order parameter saturate to its maximum value, the more robust the network is considered. That is because the maximum value corresponds to the state of no damage.

\begin{figure}[t]

\begin{minipage}{\linewidth}
\minipage{0.49\textwidth}
  \includegraphics[width=0.99\textwidth]{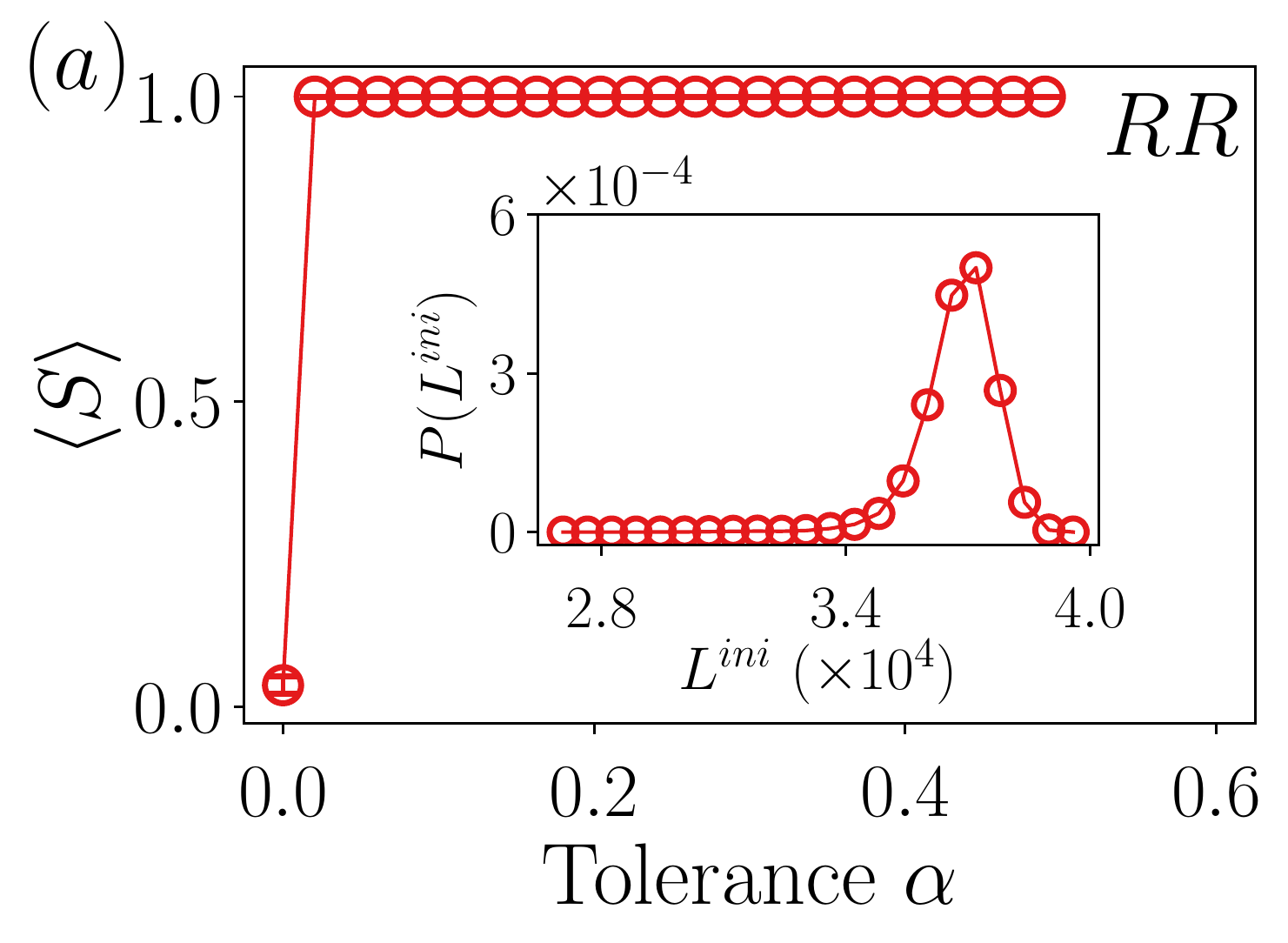}
\endminipage\hfill
\minipage{0.49\textwidth}
  \includegraphics[width=0.99\textwidth]{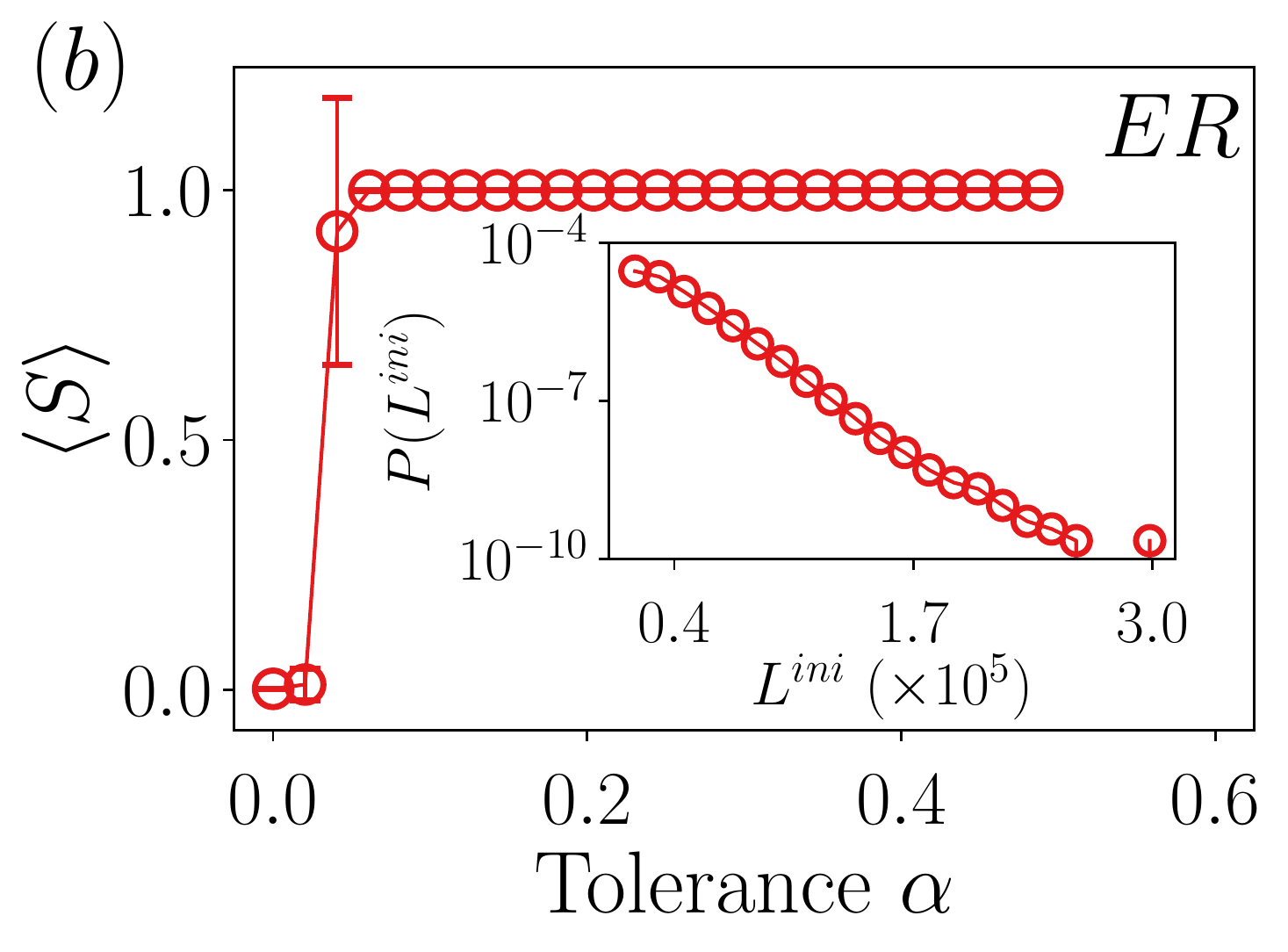}
\endminipage
\end{minipage}

\caption{Size of the largest connected component once the cascade propagation stops, for duplexes of random regular graphs~($a$) and Erd\H{o}s-R\'enyi networks ($b$). In ($a$) all nodes have the same degree $k_c = 4$, while in ($b$) the mean degree is $\langle k \rangle = 4$ and a minimum degree $k_0 = 2$ has been imposed to guarantee global connectivity. In both cases the number of nodes per layer is $N = 5000$. Error bars correspond to $1$ standard deviation, although in most of the points the error it is smaller than the marker size. Averages are taken over $1000$ realizations of the dynamics. In the insets, the probability density of the initial load $L^{ini}$ computed from $100$ independent networks.}
\label{fig:fig0}
\end{figure}

\begin{figure}[t]
  \includegraphics[width=0.33\linewidth]{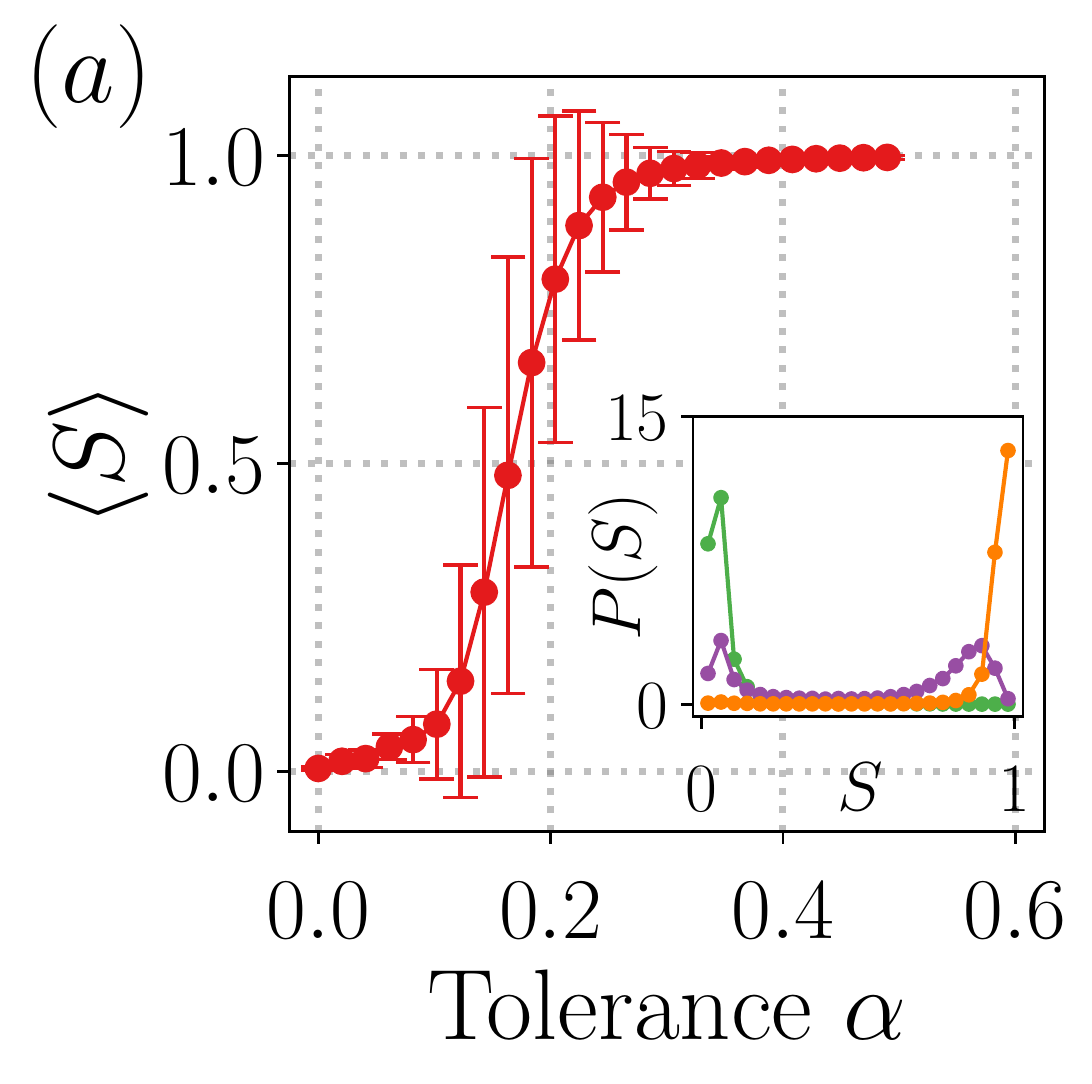}
  \includegraphics[width=0.33\linewidth]{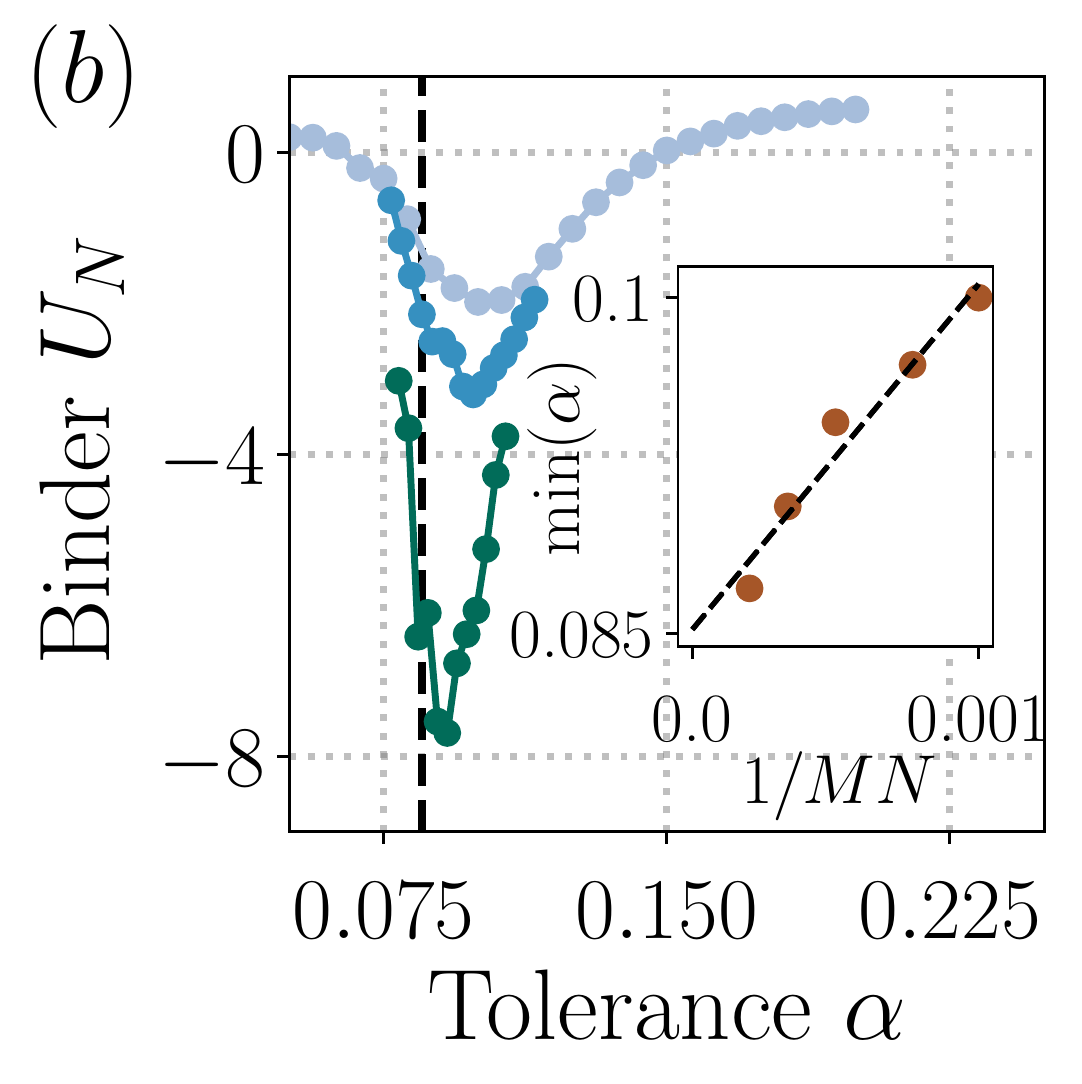}
  
  \includegraphics[width=0.55\linewidth]{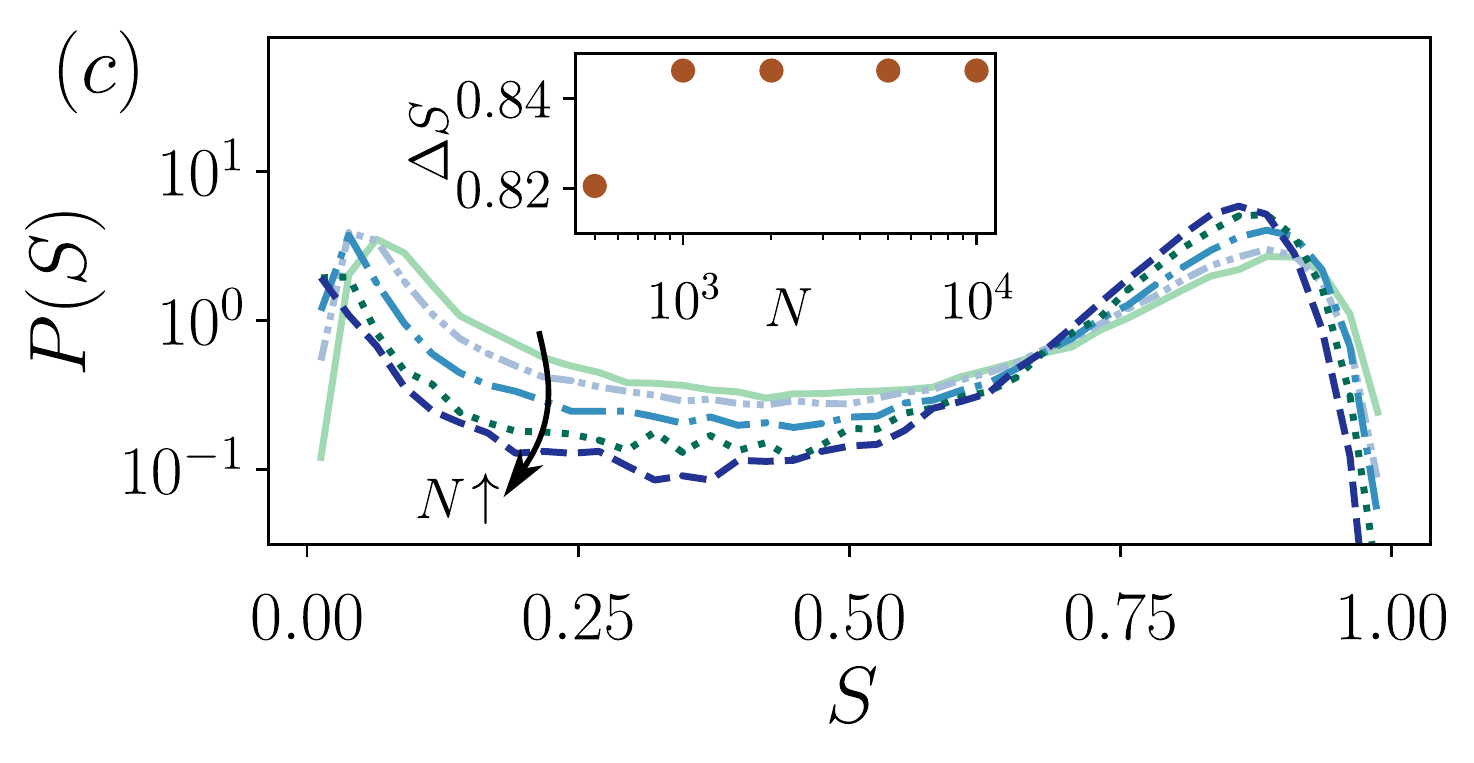}

\caption{Characterization of the abrupt transition for a duplex composed by scale-free networks (degree distribution $p(k) \propto k^{-3}$) with mean degree $\langle k \rangle = 3$, minimum degree $k_0 = 2$ and maximum degree $k_{max}\sim \mathcal{O}\left(N^{1/2}\right)$. In $(a)$, average of the size of the largest connected component as a function of the tolerance parameter. The error bars correspond to one standard deviation. In the inset, probability of $S$ for $\alpha = 0.1, \, 0.17, \, 0.25$ (see the video in the Supplemental Material for a better appreciation of the transition). The network size per layer is $N = 1000$. In $(b)$, the Binder cumulant, Eq.~\eqref{eq:binder}, for different system sizes. The dashed vertical line is located at the critical point. In the inset, position of its minima, with a linear fit (dashed line). In $(c)$, probability of $S$, from $N = 500$ to $10000$. In the inset, distance between peaks as a function of the system size. All points and histograms are computed by at least $20000$ independent realizations of the dynamics.}
\label{fig:fig1}
\end{figure}

$ $
\section{Results}

We start our analysis by studying duplexes of random regular (RR) graphs and Erd\H{o}s-R\'enyi (ER) networks, which are the simplest non-trivial network topologies. The degree distribution of the former is a Dirac delta $P(k) = \delta(k-k_c)$, i.e., all nodes share the same number of connections $k_c$, while in the latter case $P(k)$ is well-approximated by a Poisson distribution of parameter $\langle k \rangle$. In Fig.~\ref{fig:fig0} we show the effects of the cascade propagation in these types of graphs. We take as order parameter the average of the size of the largest connected component (LCC) at the end of the cascade propagation, $\langle S \rangle $. We see that, for both topologies, the destructive effects of the initial perturbation are barely appreciated for almost any tolerance $\alpha$. The only possible disruption occurs when the network is initially operating very close to its maximum capacity, i.e., when $\alpha \to 0$. The reason behind this high robustness is that nodes are statistically equivalent among them, so they all have an initial load of the same order of magnitude --- see the insets of Fig.~\ref{fig:fig0} which show that the initial load distributions are bounded and have well-defined mean and variance. When a node is attacked, be it the one with the largest load or one chosen at random, the excess of load that needs to be handled by other nodes is a small fraction of their own load. Recalling that the failure condition depends on individual initial loads, it suffices a small tolerance to avoid the cascade propagation. 

We next move our focus to cascading dynamics in a duplex of scale-free networks. This system displays a very heterogeneous load distribution~\cite{barthelemy2004betweenness}, hence we expect a different qualitative behavior in its robustness properties. Indeed, it is immediate to see [Fig.~\ref{fig:fig1}($a$)] that the transition from a completely destroyed network to an almost intact one occurs at a finite tolerance. This critical value $\alpha_c$ separates two actual regimes, or \textit{phases} in the parlance of critical phenomena: one in which the initial perturbation propagates at a network-wide scale, dismantling the network functionality, and another in which the perturbation barely affects a few nodes. Therefore scale-free multiplex networks are weaker than RR and ER multiplexes, because, when an attack occurs, the nodes need more resources (tolerance) to absorb the redistributed flow and avoid the global failure.

Knowing whether the nature phase transition is continuous or discontinuous is useful not only to shed light on the dynamics of cascade propagation but also for robustness assessments. Indeed, a discontinuous transition is much more sudden, difficult to anticipate and devastating than a continuous one. The order parameter in Fig.~\ref{fig:fig0} jumps abruptly from $0$ to $1$, thus we cannot rule out the possibility that the phase transition in multiplex scale-free networks is also discontinuous, even though the growth of $\langle S \rangle$ is much smoother in this case. The first thing to note is that the error bars of Fig.~\ref{fig:fig1} in the critical region are surprisingly large. Their origin can be understood by plotting the probability density function of the size of the LCC, $P(S)$ [see the inset of Fig.~\ref{fig:fig1}$(a)$]. For low values of the tolerance, $P(S)$ is single-peaked very close to zero. By increasing $\alpha$ we enter in the critical region and another peak develops, away from the first one. The position of these peaks is barely dependent on $\alpha$, but their relative height is. As the tolerance keeps increasing, the second peak becomes higher, until eventually the first peak disappears. This bimodality is a typical feature of abrupt phase transitions. Note, moreover, that our model is genuinely out of equilibrium, the ergodicity is broken due to an absorbing state, and the bimodality is not a signature of hysteresis or phase coexistence but reminds the absence of self-averaging~\cite{wiseman1998finite}.

To further characterize the transition we use the Binder cumulant~\cite{binder1981finite}
\begin{equation}
    \label{eq:binder}
    U_N \equiv 1 - \frac{\langle S^4 \rangle}{3\langle S^2 \rangle^2},
\end{equation}
where $\langle S^n \rangle $ is the $n$th moment of the size of the largest connected component once the cascade has stopped. The behavior of this quantity for different system sizes provides the value of the critical point and can be used to infer the nature of the phase transition. In Fig.~\ref{fig:fig1}$(b)$ we show $U_N$ for different system sizes, and we can appreciate that they display a non-monotonic behavior with a minimum that becomes lower and more narrow as $N$ increases, which is again a signature of an abrupt transition. The position of the minima can be fitted against the inverse of the system size~\cite{de2018finite}, finding $\alpha_c$ as the intercept of the fit [see the inset of Fig.~\ref{fig:fig1}$(b)$]. For the parameters used in the simulations of Fig.~\ref{fig:fig1}, we obtain $\alpha_c = 0.085 \pm 0.002$. To be sure that the phase transition remains abrupt in the limit $N \to \infty$ and is a \textit{bona fide} discontinuous transition, we verify that, at the value of the tolerance for which the susceptibility of $S$ is maximum, the value of $P(S)$ between the peaks vanishes as the size is increased [Fig.~\ref{fig:fig1}$(c)$], and that distance between the maxima $\Delta S$ does not decrease [see inset of Fig.~\ref{fig:fig1}$(c)$]~\cite{grassberger2011explosive, tian2012nature}. Notice that $\Delta S(N)  \approx 0.85$, hence the transition is devastating and very hard to anticipate. In the light of the previous findings, we devote the rest of the article to study the phase transition precisely in scale-free multilayer networks, having in mind that they represent a worst-case scenario when compared to random regular graphs and Erd\H{o}s-R\'enyi networks.

\begin{figure}[t]

\begin{minipage}{\linewidth}
\minipage{0.32\textwidth}
  \includegraphics[width=0.99\textwidth]{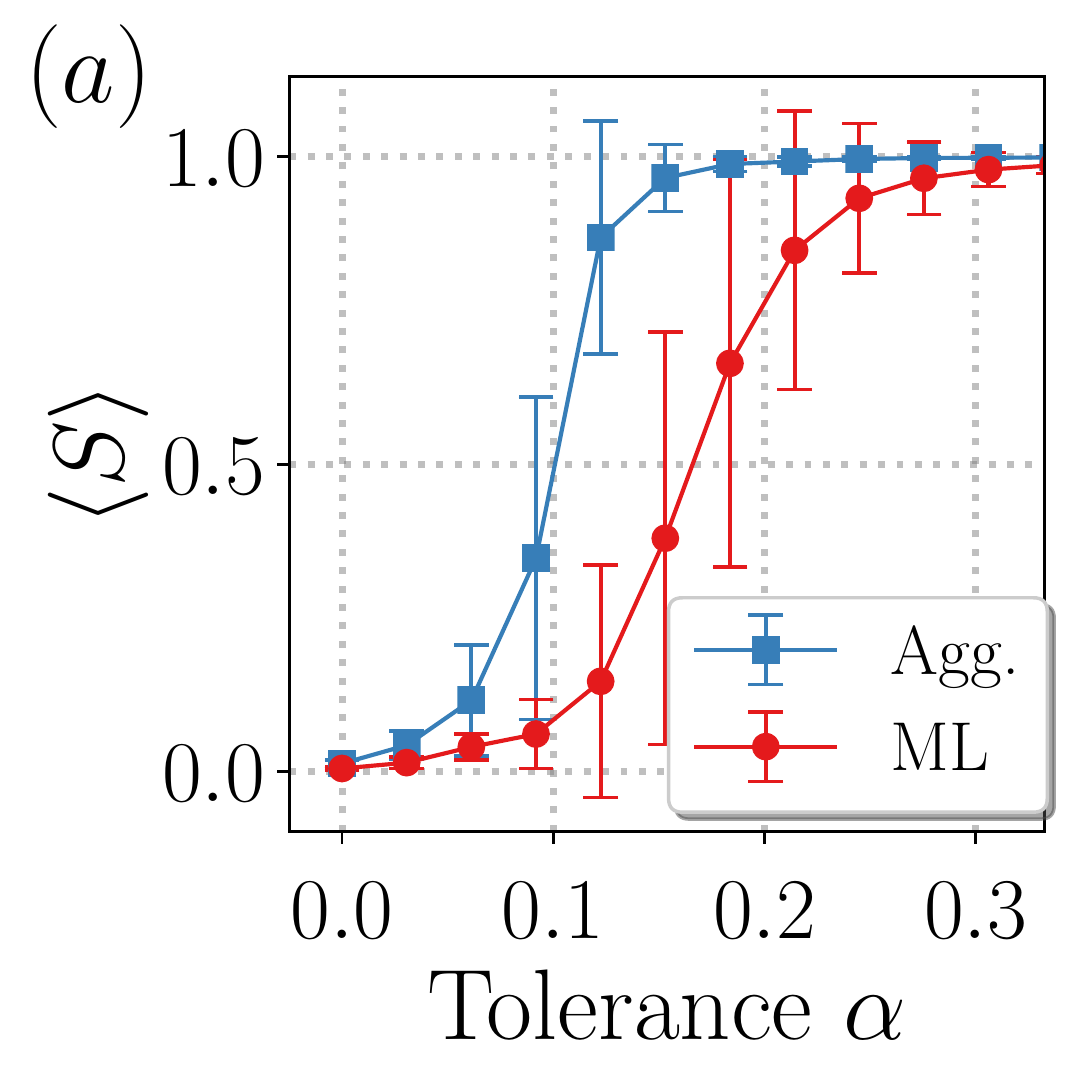}
\endminipage\hfill
\minipage{0.32\textwidth}
  \includegraphics[width=0.99\textwidth]{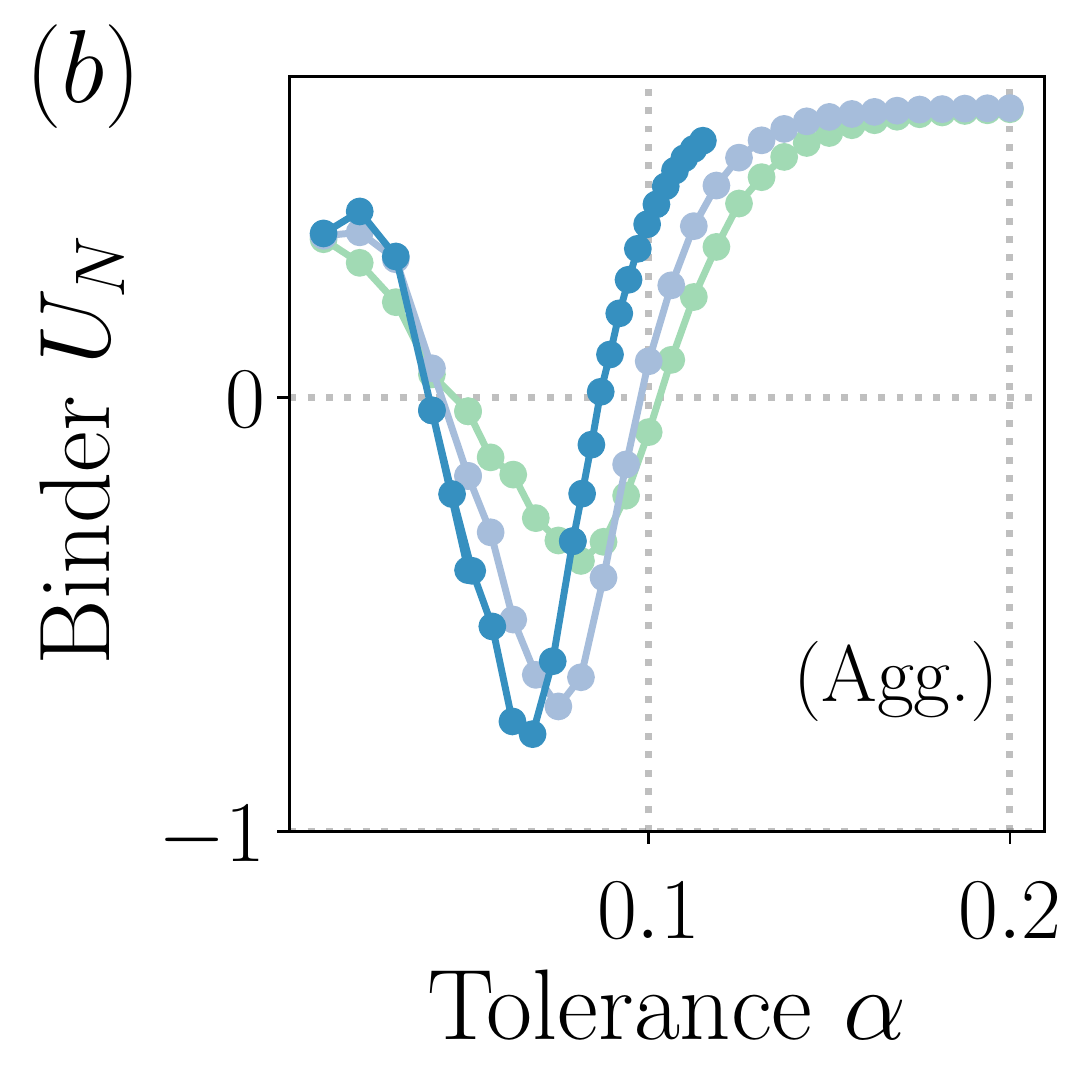}
\endminipage\hfill
\minipage{0.32\textwidth}
  \includegraphics[width=0.99\textwidth]{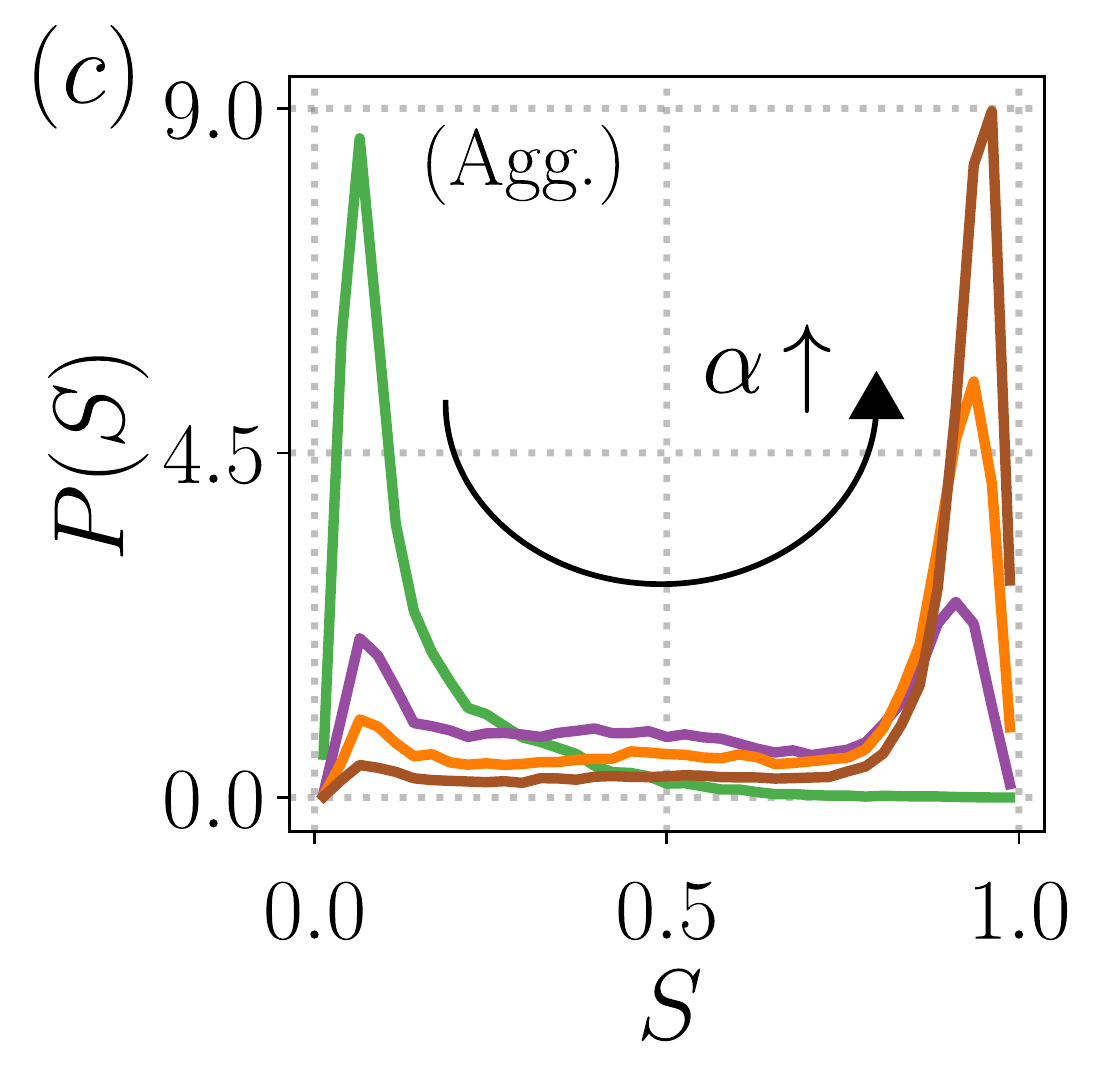}
\endminipage

\minipage{0.32\textwidth}
  \includegraphics[width=0.99\textwidth]{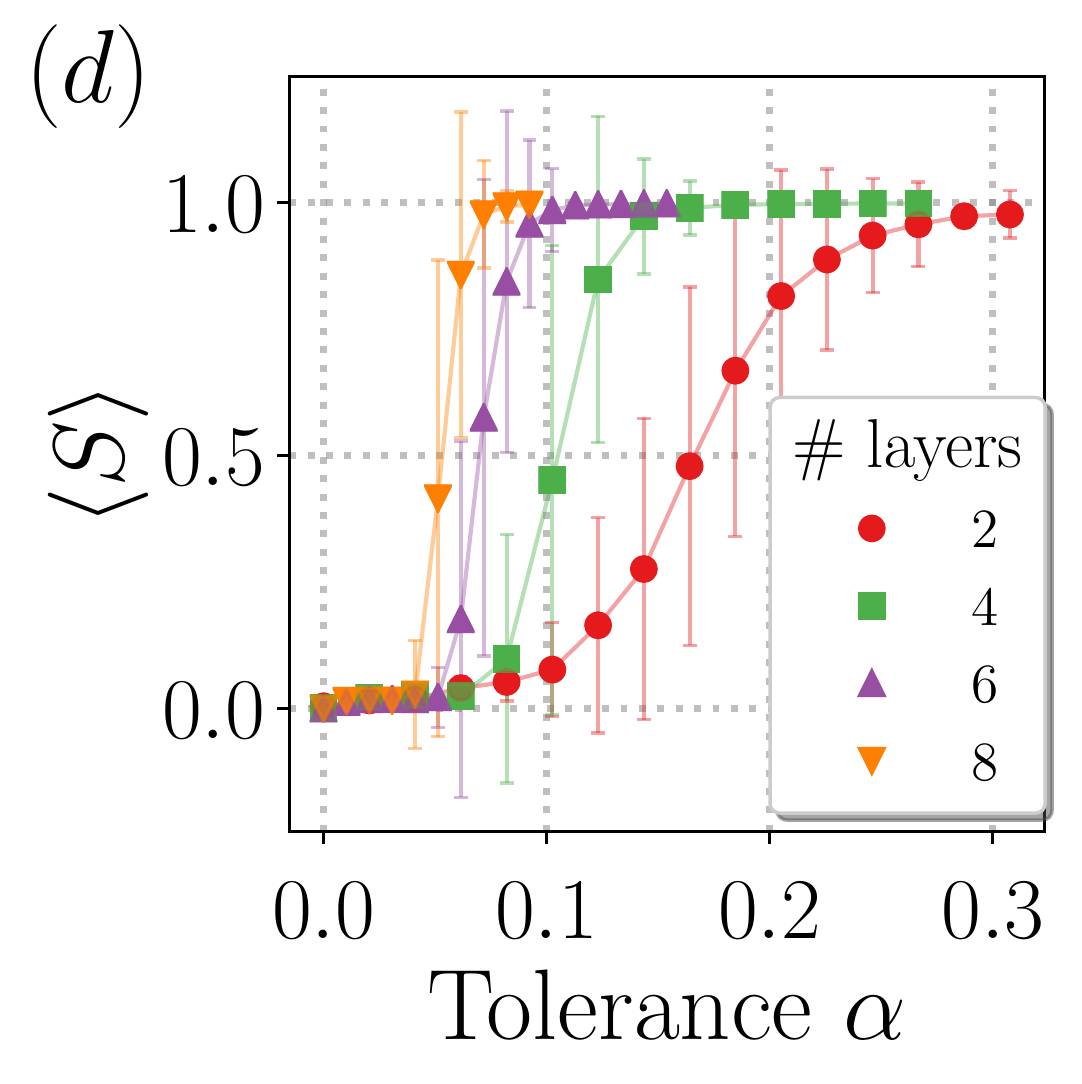}
\endminipage\hfill
\minipage{0.32\textwidth}
  \includegraphics[width=0.99\textwidth]{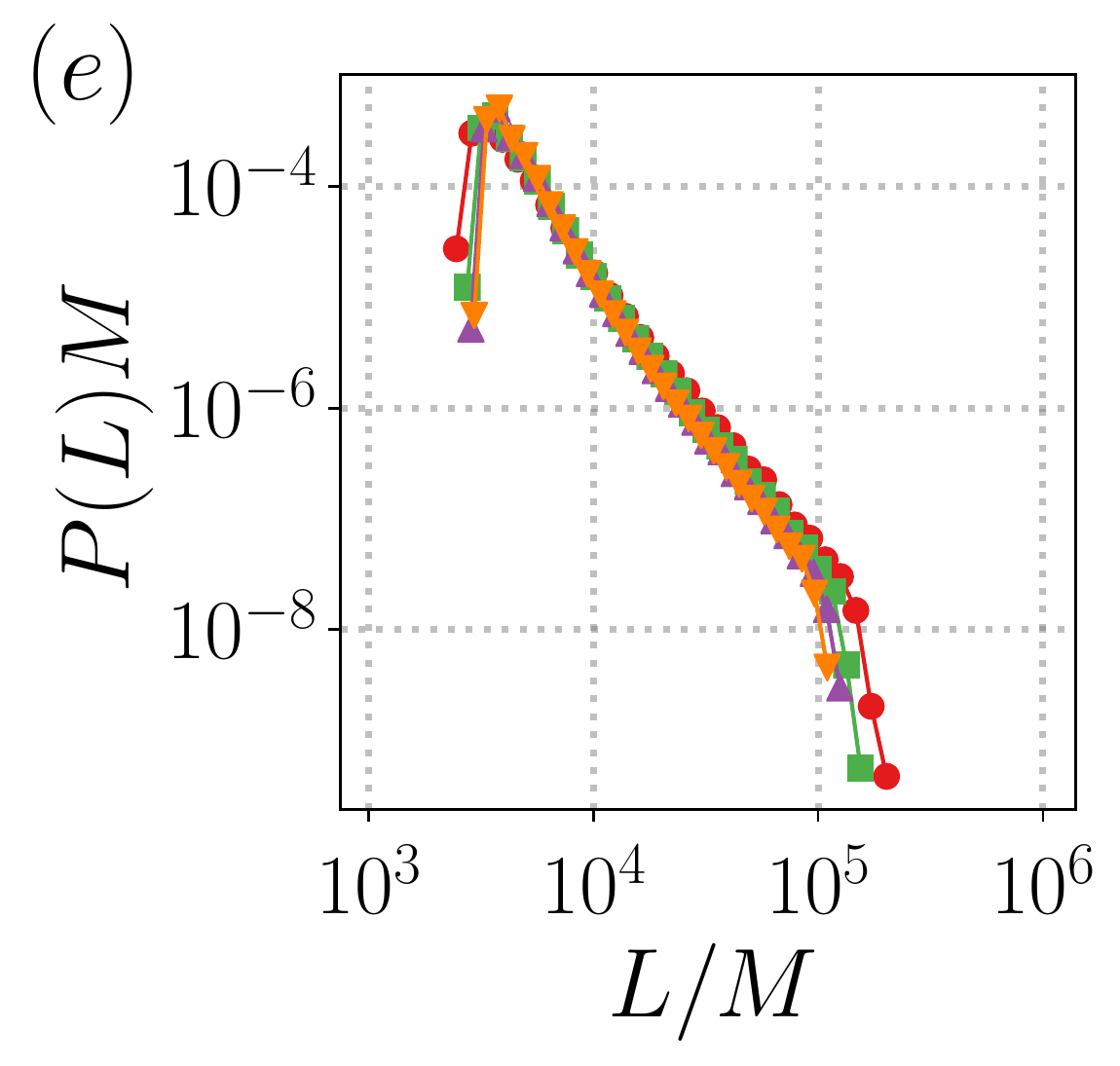}
\endminipage\hfill
\minipage{0.32\textwidth}
  \includegraphics[width=0.9\textwidth]{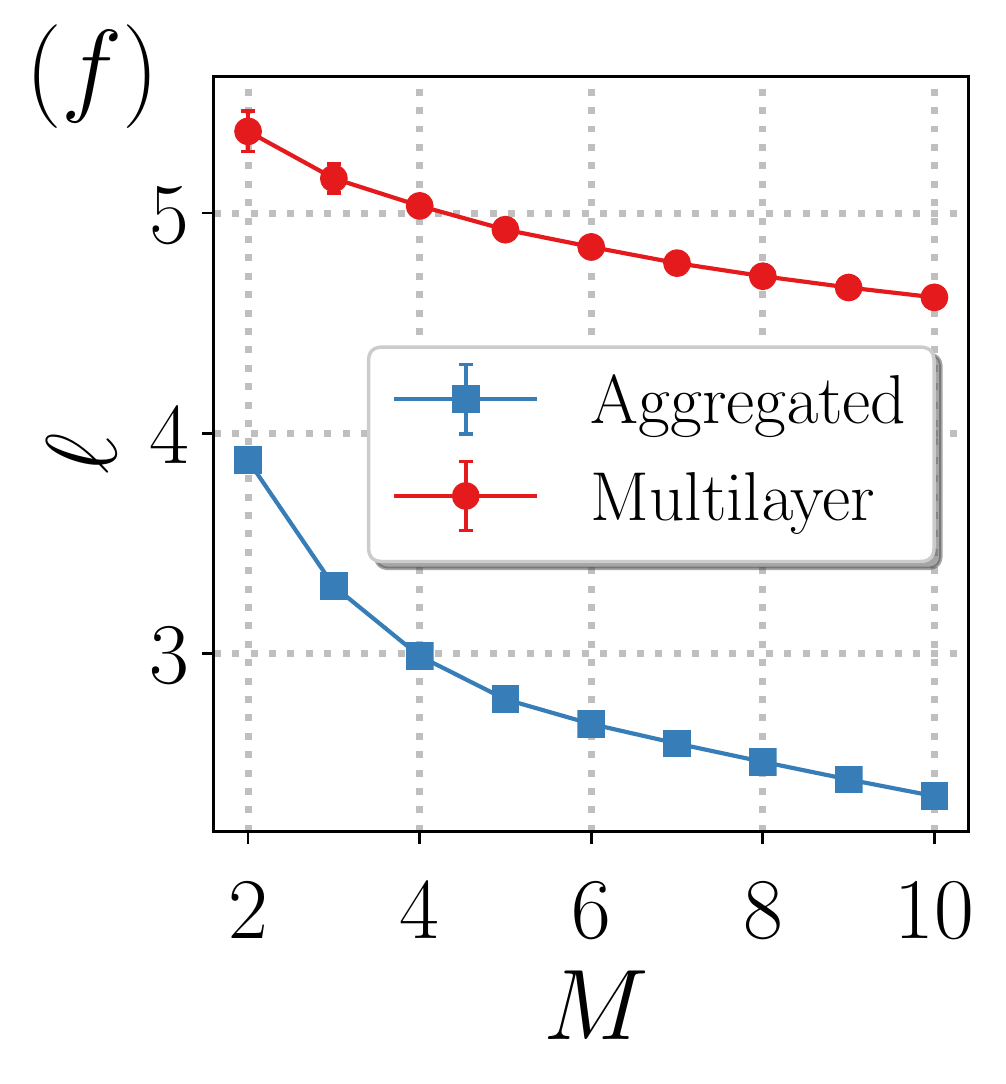}
\endminipage

\end{minipage}

\caption{In $(a)$, comparison of the robustness between a duplex ($M=2$) of scale-free networks and its aggregated counterpart. In $(b)$, Binder cumulant for aggregated networks of size $N = 250, \, 500 \text{ and } 1000$. In $(c)$, probability density function of the size of the largest connected component at the end of the cascade, for aggregated networks. Values of the tolerance are $\alpha = 0.07, \, 0.1, \, 0.11 \text{ and } 0.12$ . In $(d)$, robustness of a multiplex network as a function of its number of layers. In $(e)$, load distribution for the networks in $(d)$. In $(f)$, average path length for multiplex networks formed with different number of layers $M$ and for their aggregated counterparts. The network of each layer is constructed with the parameters indicated in the caption of Fig.~\ref{fig:fig1}. Each point in ($b$) and the histograms in ($c$) are computed from $20000$ independent realizations of the dynamics. All the other results are averaged over $1000$ realizations.}
\label{fig:fig2}
\end{figure}

The dynamics of different sort of models, not only those regarding the robustness, suffer non-trivial changes depending on whether the topology is layered or aggregated (single-layer). For example, when it comes to phase transitions, it is often observed in percolation and in first-neighbor cascade propagation that continuous transitions in monolayers become first-order when the dynamics is considered in the multilayer. This prompts us to compare the level of robustness and the type of transition for non-local cascade propagation in multiplex networks and their aggregated representation. For the aggregation process, we collapse all intralayer links into an aggregated network of $N$ nodes. For the sake of simplicity, in the aggregated structure we do not consider multiedges, although this is not really a problem as far as the number of layers of the original network is kept low, the networks are sparse and the system size is large enough, since the probability of observing an existing link in another layer scales as $\langle k \rangle / N$. The first observation is that the aggregated network is more robust than the layered one, since $\langle S \rangle$ saturates to $1$ for lower values of the tolerance [see Fig.~\ref{fig:fig2}$(a)$]. Put another way, the robustness of an aggregated network obtained from an originally layered structure will tend to be overestimated. Although the aggregated robustness is higher, the metrics to characterize the transition indicate that the system behaves as in the multiplex. Indeed the Binder cumulant becomes negative in the critical region, with a narrower and deeper minimum as the system size increases [Fig.~\ref{fig:fig2}($b$)], and the distribution of cascade sizes maintains the desired bimodality [Fig.~\ref{fig:fig2}($c$)].

The robustness overestimation effect can be understood from topological arguments. The aggregated network, by construction, is less heterogeneously connected than the multiplex one. It is well-known that degree heterogeneity amplifies the cascade propagation (see Ref.~\cite{motter2002cascade}, but also our above results) hence we can expect a higher robustness in the former. Moreover, aggregated networks are less sparse than layered ones: in the former the mean degree is $\langle k \rangle M $, while in the latter is $\langle k \rangle+M-1$. This sparsity is reflected in the average shortest path length $\ell$, which is smaller for aggregated networks: for Fig.~\ref{fig:fig2}$(a)$ we obtain $\ell_{agg} = 3.89 \pm 0.05$ and $\ell_{ML} = 5.4 \pm 0.1 $. We will see that $\ell$ correlates well with the level of robustness of the network.

Most real multilayer networks have been formed or constructed in a non-optimal way, in the sense that there exist redundant layers to the overall functioning of the network \cite{de2015structural, lacasa2018multiplex, ghavasieh2019reducing}. This is natural, as multilayer systems usually are not designed at once but their structure is a result of several growing and shrinking processes occurred at different time scales. This prompts us to ask how the robustness of a layered system depends on the number of layers. We see in Fig.~\ref{fig:fig2}$(d)$ that adding more layers causes an increase of robustness. This effect is not rooted in a change of the load heterogeneity due to the addition of layers: see the collapse of the load distributions in Fig.~\ref{fig:fig2}$(e)$. Instead, the global connectivity is enhanced as $M$ increases, so the load excess caused by a perturbation can be absorbed more easily by the rest of the network. This is reflected in the decreasing tendency of the average path length, see Fig.~\ref{fig:fig2}$(f)$. An apparent paradox is posed: why both collapsing the layers into an aggregated network and unfolding a multiplex network in an even more layered structure turn the system more robust? The answer is simply that these two procedures generate less sparse networks, i.e., it is observed a decrease in the average path length (compare the two curves in Fig.~\ref{fig:fig2}$(f)$). Note, however, that here we are dealing with synthetic multiplex networks, with statistically similar topologies across layers: same number of nodes $N$ and same degree distribution $P(k)$. Moreover, they have been constructed with no structural correlations, neither at the intralayer level nor between layers. Therefore, the further we are from this setting, the more uncontrollable effects might occur in the processes of aggregation or addition of layers and the less reliable $\ell$ might become to assess the robustness.

\begin{figure}[t]
\begin{minipage}{\linewidth}
\minipage{0.45\textwidth}
  \includegraphics[width=0.99\textwidth]{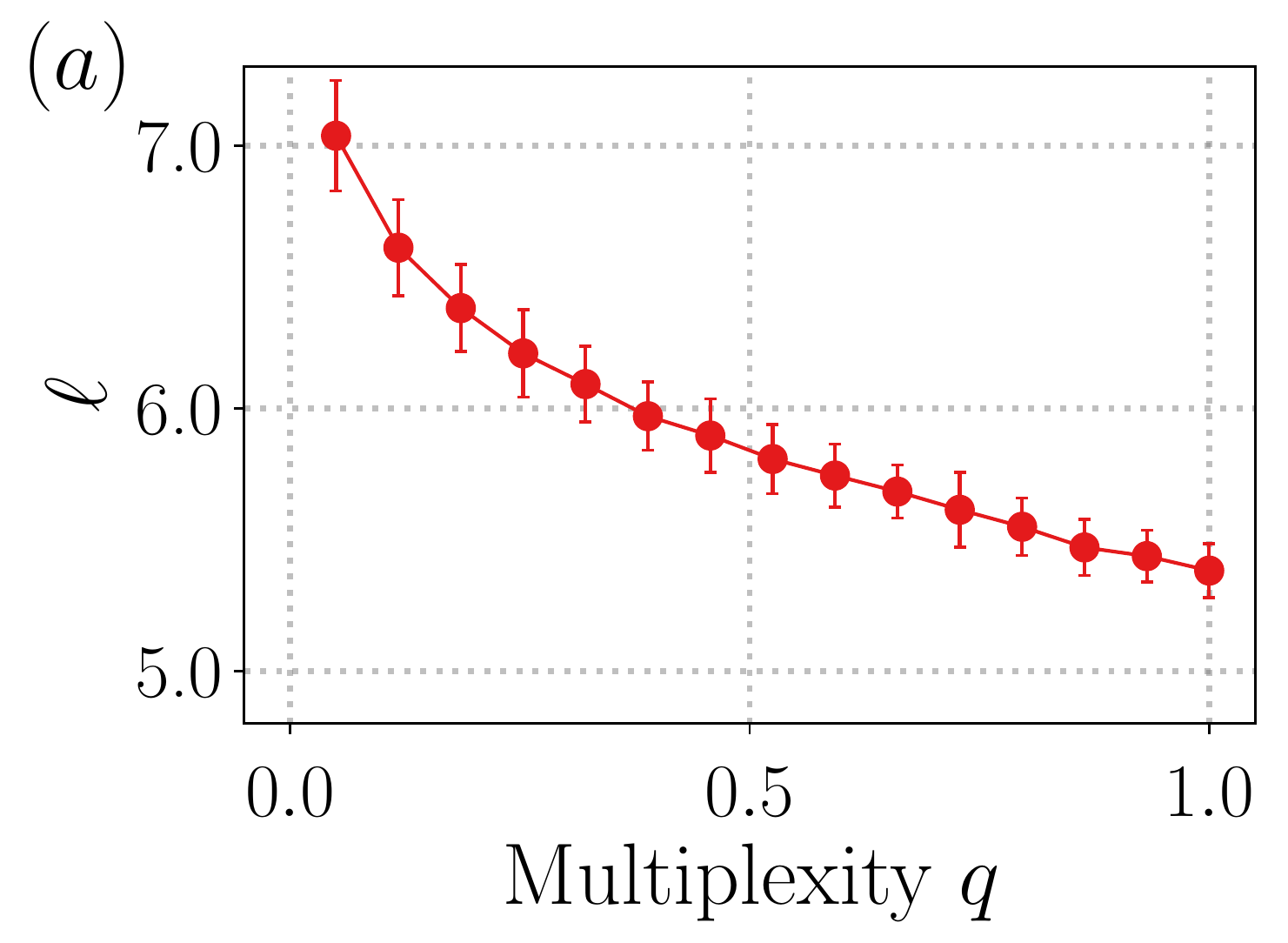}
\endminipage\hfill
\minipage{0.45\textwidth}
  \includegraphics[width=0.99\textwidth]{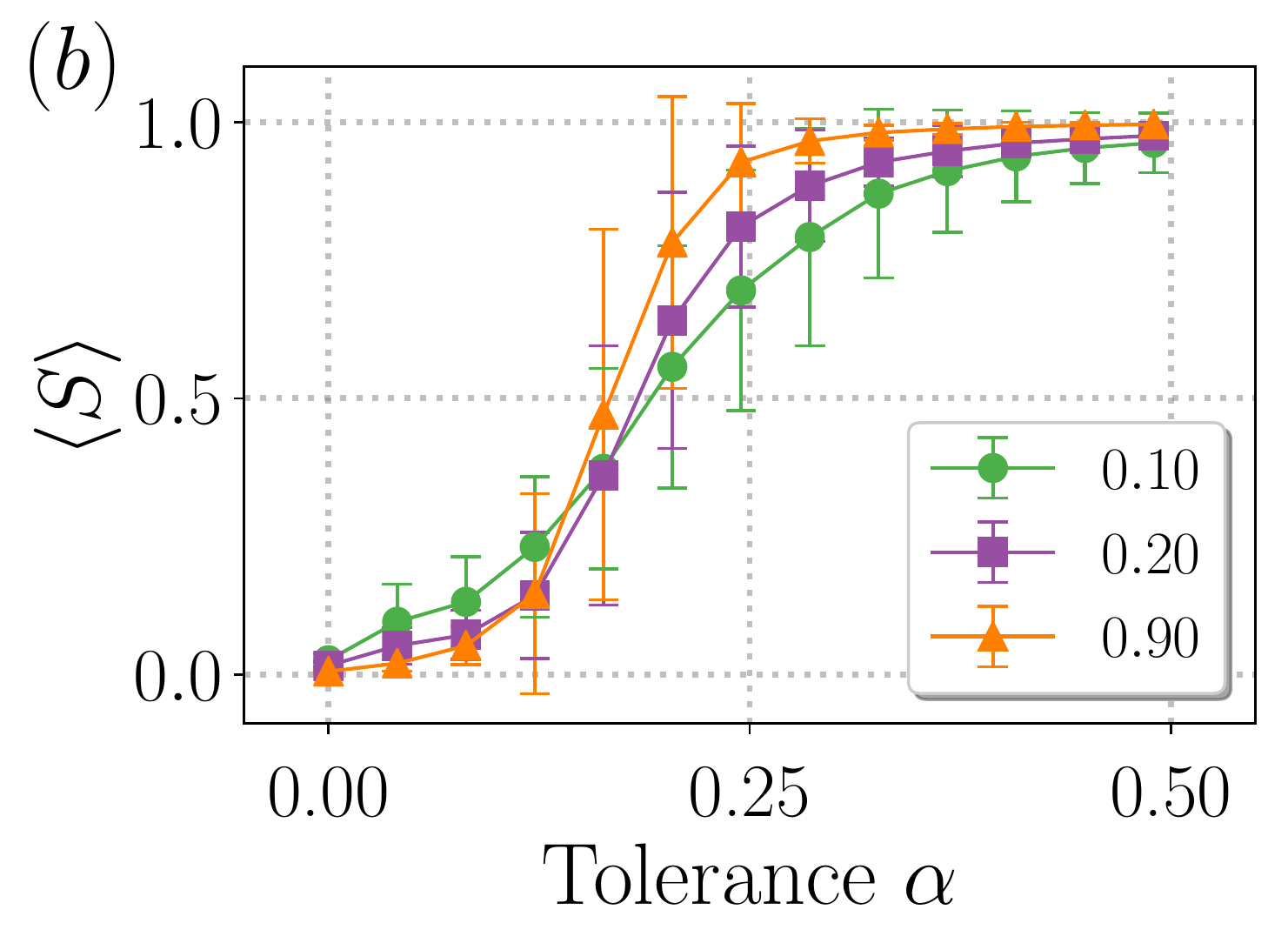}
\endminipage
\end{minipage}

\caption{In $(a)$, average path length as a function of the multiplexity parameter. In $(b)$, average of the size of the largest connected component for different values $q$, indicated in the inset. Each point is averaged over $1000$ independent realizations of the dynamics, in networks with the same properties as those of Fig.~\ref{fig:fig1}.}
\label{fig:fig3}
\end{figure}

To provide further evidence of the relation between the average path length and the robustness of the system, we next study the case of partial multiplexity, which is another feature present in real networks and is known, as well, to strongly modify the dynamics of models running on multilayer networks~\cite{artime2017joint, diakonova2016irreducibility}. Let us focus in the scenario of two layers, for simplicity. In this case, the multiplexity parameter $q$ is the fraction of nodes that simultaneously participate in both layers, that is, they share the same state ---functioning or failed--- and are connected via an interlayer link, hence contributing to the load. This way, when $q$ is low, the common nodes in the multilayer are highly loaded since they act as a bridge between the two layers. At the global level, higher multiplexity implies lower average path length, as it can be seen in Fig.~\ref{fig:fig3}$(a)$. Thus, for a fixed and finite $N$, we expect systems with large $q$ to be more robust than those with a lower value. Indeed, as displayed in Fig.~\ref{fig:fig3}$(b)$, this is the behavior that we find: the value of the tolerance such that a cascade has barely affected the network ($S \sim 1$) for $q = 0.1$ is almost the double of the one for $q = 0.9$. 

Up to this point we have considered statistically similar topologies drawn from the configurational model. However, in most social, biological and technical networks this homogenization across layers does not exist, and intralayer and interlayer correlations are present, so we next evaluate the propagation of nonlocal cascades in a real multilayer system. We take the network of airline companies of the European air transportation network \cite{cardillo2013emergence}. Each layer corresponds to a company, and there are a total of $37$. Nodes are the airports and intralayer links are present if a company has operating flights between those airports. The load on this network can be seen as the flow of passengers, and interlayer links can be interpreted as passenger layovers.

\begin{figure}[t]
\begin{minipage}{\linewidth}
\minipage{0.49\textwidth}
  \includegraphics[width=0.99\textwidth]{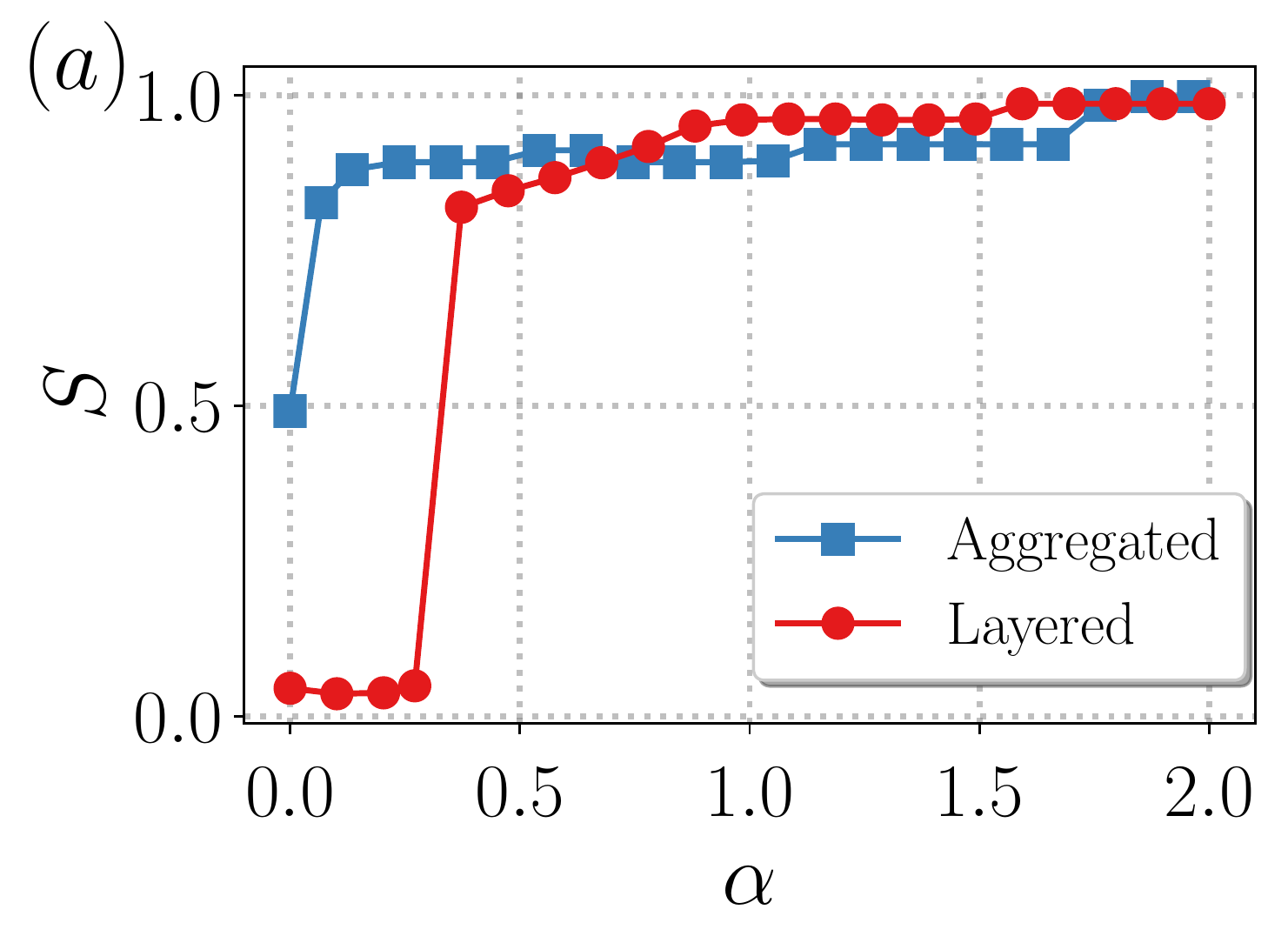}
\endminipage\hfill
\minipage{0.49\textwidth}
  \includegraphics[width=0.99\textwidth]{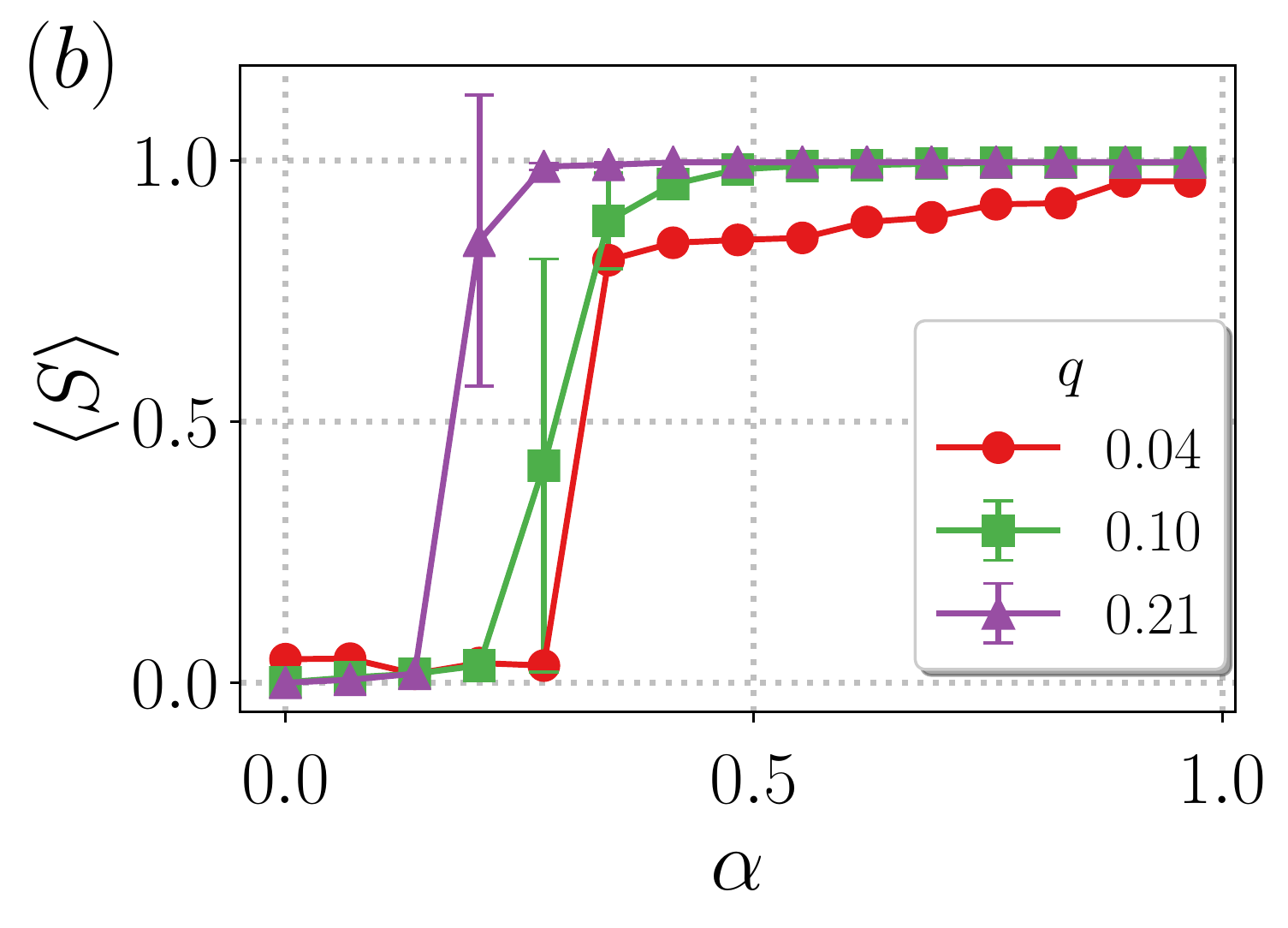}
\endminipage
\end{minipage}

\caption{In $(a)$, comparison of the robustness of the layered and aggregated versions of the European air transportation network. In $(b)$, size of the LCC as the multiplexity is increased with the method explained in the main text. The curve for $q = 0.04$ corresponds to the original layered networks, without extra interlayer links. Each point is averaged over $10$ independent realizations of the dynamics.}
\label{fig:fig4}
\end{figure}

Simulating the propagation of cascades in the air transportation multiplex and its aggregated version, we find that, as anticipated, the process of aggregation increases the robustness [see Fig.~\ref{fig:fig4}$(a)$]. Moreover, the transition in the layered version of the network occurs at a finite tolerance and the jump is quite significant, agreeing with the expected result of a discontinuous transition.

We also look at the role played by the multiplex parameter $q$. Recall that for a duplex, $q$ represents the number of nodes participating simultaneously in the dynamics and sharing the same state. When $M > 2$, however, new possibilities unfold, since a node might be present in some layers but absent in other ones. Indeed, this occurs in our example of the European air transportation network: for a given airport, there are some companies that operate there and others that do not, although technically they would be allowed to do so. We can generalize $q$ in order to include this scenario as follows. Let $n_i$ be the number of times node $i$ is present (with intralayer neighbors) in a multiplex network. In our case, this is the number of airlines operating in a given airport. The actual number of interlayer links between that node and its replicas is then $n_i(n_i -1)/2$. Let $n_T$ be the total number of possible interlayer links of a node, even if it does not have intralayer connections, i.e., $n_T = M (M-1)/2$. Then the multiplexity parameter for such a network is defined as the ratio between the actual number of interlayer links and all the possible interlayer links, that is,
\begin{equation}
    q = \frac{1}{N M(M-1)} \sum_{i=1}^{N} n_i (n_i - 1).
\end{equation}
Note that the duplex case is included in this new definition.

We can increase the $q$ of a multiplex network by taking an inactive node of a layer (e.g, an airport in which a company does not operate) and creating some intralayer connection(s) to it. This way that node becomes active in the layer, permitting interlayer flow (e.g., passengers can take a layover from another layer and then fly using that company). Although many strategies can be devised to increase $q$ taking into account topological considerations, we do so in the simplest possible manner: we randomly pick a node with no neighbors in a random layer and connect it to a node with neighbors on that layer, randomly selected as well. The more nodes we add, the higher will be the multiplexity $q$. In Fig.~\ref{fig:fig4}$(b)$ we confirm that the air transportation network becomes more robust as $q$ is raised, coinciding with the results obtained in synthetic networks. Certainly the robustness can be even further increased by applying strategies other than random, but this is out of the scope of the present article.

\section{Conclusions}
In this article we have tackled the problem of cascades of failures that propagate in multiplex network in a non-local way. That is, the failure of a node can be induced by the failure of other nodes not necessarily linked to it. To this goal we define the rules of a load-capacity model in multilayer topologies. In such a model, nodes are assigned a variable load and a time-independent capacity, and they fail when, for some reason, their load exceeds the capacity, i.e., we encounter cascades of overload entities. Due to the non-local dynamics and the non-trivial topologies on which the dynamics runs, the computational approach to the problem is more accessible than an analytical treatment, thus we offer a numerical characterization of the phenomena. The robustness of a network is related to purely structural properties, i.e., to what is the largest portion of the network that remains connected once the cascade stops.

We have given compelling evidence of an abrupt phase transition, rooted in the intrinsic dynamics of non-local propagation of the cascades. Hence, we have confirmed that non-local flow redistribution in multiplex networks shares the same fragility as the one observed in local spreading phenomena in these topologies, and, additionally, we have shown that this qualitative similarity between processes is broken for single-layer networks, being the non-local failure transition discontinuous too. To the best of our knowledge, this is the first time that such transition is characterized for a load-capacity model in layered systems. We observe that this transition has a dependence on the network topology. To shed some light on this relation, we work with the hypothesis that the amount of load that needs to be redistributed once a node fails is more evenly distributed across the network if the average shortest distance between nodes is small. We check this hypothesis via three independent processes that induce a decrease in the average path length: the collapse of a layered topology into a single layer, the addition of new layers, and the increase of the common nodes participating simultaneously in the layers. In all cases, we find that the robustness is increased. Finally, we have tested our predictions on the European air transportation multilayer network. There we have shown that the transition is discontinuous, and that the aggregate version of the network overestimates the real robustness of the original system. We have also defined a way to generalize the multiplexity parameter $q$ to more than 2 layers, and have shown that when it increases, the robustness does it as well.

Our work is a first approximation to the study of non-local failure propagation processes in multiplex structures. As such, we have identified the interesting phenomenology that one can find, i.e., an abrupt transition, and we have studied how to increase or decrease the robustness of the network by tuning a topological variable such as the average path length. However, several interesting questions and extensions have been left out of this first characterization of overload cascades on multiplexes, although we believe that our findings can serve to pave the way for further research on this topic. Some of them are a better description of the phase transition in terms of the load distribution, what is the impact of link weights on the critical properties, and a detailed analysis of the robustness for statistically dissimilar networks in the different layers, as well as including structural correlations both in the intralayer and in the interlayer sense. Note also that an underlying assumption of the model is that nodes communicate among them via the shortest path, which is an approximation that might fail in some cases. Hence studying the same dynamics of load redistribution where the load is computed with other centrality measures is certainly an interesting extension for the future. The overarching goal would be to develop an analytical theory of non-local propagation of overload cascades that is able to support the results found computationally and flexible enough to encompass the open questions mentioned above.

\bibliography{biblio.bib}

\end{document}